\documentclass{iopart}
\newcommand{\pdif}[2]{\frac{\partial #1}{\partial #2 } }
\newcommand{\diff}[2]{\frac{\mbox{d} #1}{\mbox{d} #2 } }

\begin{document}
\paper[MCT and the fluctuation-dissipation theorem]
{Mode-coupling theory and the fluctuation-dissipation theorem
for nonlinear Langevin equations with multiplicative noise} 
\author{Kunimasa Miyazaki and David R. Reichman}
\address{Department of Chemistry,
Columbia University, 3000 Broadway, New York, NY 10027, U.S.A}

\ead{km2239@columbia.edu}
\date{\today}

\begin{abstract} 
In this letter, we develop a mode-coupling theory for a class of 
nonlinear Langevin equations with multiplicative noise 
using a field theoretic formalism. 
These equations are simplified models of realistic colloidal
 suspensions.  
We prove that the derived equations are consistent with the
 fluctuation-dissipation theorem.  
We also discuss the generalization of the result given here to real
 fluids, and the possible description of supercooled fluids in the aging
 regime. 
 We demonstrate that the standard idealized mode-coupling theory is not
 consistent with 
the FDT 
in a strict field theoretic sense.
\end{abstract}

\section{Introduction}

Mode-coupling theory (MCT) has been a useful first principles 
approach for studying the dynamics of supercooled liquids (see, for 
example, Ref.\cite{gotze1992}).  
MCT was originally derived using projection operator methods 
together with several uncontrolled approximations. 
The theory has been successful in providing a quantitative account of
many experimental and numerical observations, such as  
nonergodic parameters and structural relaxation
exponents\cite{gotze1999}.  
The approximate nature of the conventional MCT, however, has restricted  
its validity to description of only (i) two-point correlation
functions\footnote{See, however, G. Biroli and
J. -P. Bouchaud\cite{biroli2004}. 
Note that, strictly speaking, the mode-coupling approach taken here is
of the field-theoretic, diagrammatic variety. This is what allows the
authors to compute multi-point correlation functions, something that is 
difficult to do in a useful manner from the projection operator
approach. The field theory used in the work of Biroli and Bouchaud is
that of Das and Mazenko\cite{das1986} which assumes the FDT from the
start. The work presented here is a first step beyond such a simplified
treatment.},    
(ii) for systems at equilibrium, 
and (iii) at relatively high temperatures below which
the theory predicts a spurious glass transition. 

Experimental and numerical studies, on the other hand, have 
provided us with a rich dynamics, none
of which conventional 
MCT can explain.  
For example, 
simulations and experiments of supercooled fluids 
have revealed the existence of the correlated {\it local} dynamical
heterogeneities(see, for example, Ref.\cite{ediger2000}), and dynamical
scaling\cite{yamamoto1998c}.  
Another example is supercooled liquids brought out of equilibrium
by quenching the system to low temperature or by adding shear flow.  
Here many experiments and simulations have shown violation of 
the fluctuation-dissipation theorem (FDT) and the existence of 
the effective temperatures\cite{kob2000b,berthier2002b,crisanti2003}. 
A microscopic theory which goes beyond MCT is desirable to account
these phenomena. 
For such purposes, a systematic field theoretic approach is 
a good candidate. 
Such a field theoretical perturbation scheme for dynamical processes
has been developed by Martin, Siggia, and Rose (MSR)\cite{martin1973}. 
This approach is advantageous over the projection operator technique
in that (i) it is conceptually straightforward to extend 
MCT-type equation to incorporate higher order moments of fluctuations, 
(ii) it is useful for calculating multipoint correlation functions
which are essential observables to monitor the dynamical heterogeneities
in the supercooled fluids, 
and  
(iii) it enables one to treat 
the correlation function and the response function (which are
related by the FDT if equilibrium holds) on an equal footing and,
therefore, it is a powerful tool for the treatment of nonequilibrium
systems.   

Conventional MCT used for glassy or disordered systems
is believed to be equivalent with renormalized perturbation
theory\cite{McComb2004} without vertex corrections within the 
standard loop expansion of the MSR formalism\cite{martin1973}. 
This is true for a certain class of disordered systems
such as the $p$-spin spin glass models\cite{bouchaud1996}. 
For supercooled fluids, however, this field theory has
never been systematically used to derive the MCT equation {\it even at
equilibrium}.  
The primary obstacle is that, although the original equations of motion 
satisfy the FDT at equilibrium, this does not imply  that an arbitrary
perturbation scheme also preserves the constraints of the FDT at
each order of the expansion. 
Derivations of MCT for supercooled fluids 
from the field theoretic point of view has been discussed 
by several authors\cite{das1986,schmitz1993,kawasaki1997e} but 
either the FDT has been assumed (rather than consistently
derived)\cite{das1986,kawasaki1997e}  or 
a certain model has to be introduced for the derived
equation to guarantee the FDT\cite{schmitz1993}.  
Difficulties in deriving the MCT equations and extending them to 
higher order by a systematic loop expansion for 
supercooled fluids are
due to certain properties of  
the nonlinearities in the microscopic Langevin equation. 
These difficulties do not exist in the schematic $p$-spin models. 
To illustrate the difficulties, let us consider a Langevin equation
which describes dynamics of the density field, $\rho({\bf r}, t)$, of  
the dense colloidal suspension, as an example\cite{dean1996}; 
\begin{equation}
\fl\hspace*{1.0cm}
\pdif{\rho({\bf r},t)}{t}
= D\nabla\cdot\left\{ \nabla \rho ({\bf r},t)
-\rho({\bf r},t)\nabla \int\!\!{\mbox{d}}{\bf r}'~
c({\bf r}-{\bf r}')\delta\rho({\bf r}',t)
\right\} + f_{\rho}({\bf r}, t)
, 
\label{eq:dean}
\end{equation} 
where $D$ is the diffusion coefficient which is assumed to be a
constant, $\delta\rho({\bf r},t) = \rho({\bf r}, t) - \rho_{0}$ with
$\rho_{0}=\langle \rho({\bf r},t)\rangle$ is a density fluctuation, 
$f_{\rho}({\bf r}, t)$ is a random noise, 
and $c({\bf r})$ is the direct
correlation function. 
The second term in the brackets accounts for the interaction between
the particles and is the source of the nonlinearity of the Langevin
equation\footnote{Eq.(\ref{eq:dean}) is already
coarse-grained in a sense that the bare interaction potential
is replaced with the effective potential 
$-c({\bf r})/k_{\mbox{\scriptsize B}} T$.
See Refs.\cite{kawasaki1994,Archer2004}}. 
Eq.(\ref{eq:dean}) is known to exhibit the glassy properties at high
densities.  
It has been also shown 
that, in equilibrium, this equation can be reduced to 
a standard MCT equation in the overdamped limit, using 
projection operators\cite{cichocki1987,szamel1991,franosch1997}. 
In order to see the difficulty in applying the field-theoretic MSR, let
us rewrite eq.(\ref{eq:dean})  
in a following form;
\begin{equation}
\pdif{\rho({\bf r},t)}{t}
= \int\!\!{\mbox{d}}{\bf r}'~
L_{\rho({\bf r})\rho({\bf r}')}
\frac{\delta S}{\delta \rho({\bf r}')}
 + f_{\rho}({\bf r}, t), 
\label{eq:dean2}
\end{equation} 
where
\begin{equation}
L_{\rho({\bf r})\rho({\bf r}')}
={k_{\mbox{\scriptsize B}}}^{-1}D\nabla\cdot
\nabla'\rho({\bf r},t) \delta({\bf r}-{\bf r}')
\label{eq:Lij}
\end{equation}
is the Onsager coefficient and 
$S$ is the entropy of the whole system which is given as 
a functional of the density
by 
\begin{equation}
\fl\hspace*{0.5cm}
S = k_{\mbox{\scriptsize B}}\left\{
-\int\!{\mbox{d}}{\bf r}~\rho({\bf r})
\left[ \ln\left\{\rho({\bf r})/\rho_{0} \right\}-1\right]
+\frac{1}{2}\int\!{\mbox{d}}{\bf r}\int\!{\mbox{d}}{\bf r}'~
c(|{\bf r}-{\bf r}'|)\delta\rho({\bf r})\delta\rho({\bf r}')
\right\}.
\label{eq:S0}
\end{equation}
The random field $f_{\rho}({\bf r}, t)$ satisfies 
\begin{equation}
\left\langle 
f_{\rho}({\bf r}, t)f_{\rho}({\bf r'}, t')
\right\rangle
= 2{k_{\mbox{\scriptsize B}}}L_{\rho({\bf r})\rho({\bf r}')}\delta(t-t').
\label{eq:2ndFDT}
\end{equation}
One sees that there are two types of nonlinearities entangled in 
eq.(\ref{eq:dean2}); One is due to the 
thermodynamic force $\delta S/\delta \rho({\bf r})$, which is nonlinear 
in $\delta\rho$. 
Another is the density dependent 
Onsager coefficient which, through eq.(\ref{eq:2ndFDT}), makes the 
random force a nonlinear function of the density, producing
{\it multiplicative noise}. 
These properties are quite general for Langevin equations for 
realistic fluids.
As elucidated in the next section, 
these two nonlinearities are the origin of difficulties
which hamper the field theoretical approach for realistic fluids. 

In this letter, we develop the tools to treat some of the difficult
nonlinearities discussed above, including multiplicative noise.
Our goal is to derive MCT-type equations which satisfies
the FDT at the lowest level of the loop expansion 
if the system is at equilibrium. 
This goal is an
important prerequisite condition when we set out to explore 
nonequilibrium systems. 
Our theory will serve as a first step, however incremental, 
to prepare for the development of a field theory for supercooled fluids 
which goes beyond the conventional MCT and 
is capable of exploring, for example, 
nonequilibrium systems, 
the effect of higher order loops, 
and multipoint correlation functions. 
In particular, we will show that:
\vspace*{0.1cm}

\noindent
(a) Care must be exercised in field theoretical derivations of the
response of a system described by eq.(\ref{eq:dean}) to an external
field since the response function is not trivially connected to the
propagator in general. 
This fact has been ignored in past works\cite{deker1975}. 

\vspace*{0.1cm}

\noindent
(b) The multiplicative noise term is essential for a proper treatment of
the memory term. 
This point has been overlooked in past works\cite{kawasaki1997e}, as the
proper field theoretic treatment of this term is subtle. 

\vspace*{0.1cm}

\noindent
(c) A standard one-loop treatment can lead to the usual MCT of
G{\"{o}}tze and coworkers\cite{gotze1992,gotze1999} at the expense of
satisfying the FDT. 
This will lead us to consider a slightly simpler model for which the
associated self-consistent one-loop theory presents the FDT. 
This model will allow us to make connection with the work of Schmitz
{\it et al.}\cite{schmitz1993}. 

\vspace*{0.1cm}

\noindent
In the next section, we give more detailed accounts of the background  
and motivation of the present work. 
Section 3 is devoted to derivation of MCT for
multiplicative noise using the MSR method. 
Consistency with the FDT is discussed
in Section 4 and in Section 5. 
Perspectives for developing MCT for the nonequilibrium case are
discussed in Section 6.

\section{Background}

In order to make the argument general, let us consider a classical
stochastic dynamical process of a field variable 
$x_{i}(t)$, where $i$ is an index that denotes the type of field
(such as the density) and coordinates (such as positions or wave 
vectors) which may be either discrete or continuous. 
If the system is macroscopically at equilibrium, $x_{i}(t)$ obeys a
nonlinear Langevin equation of a general form:
\begin{equation}
\dot{x}_{i}
= M_{i\alpha}\pdif{S}{x_{\alpha}} 
+ L_{i\alpha}\pdif{S}{x_{\alpha}} + f_{i}
\equiv K_{i\alpha}\pdif{S}{x_{\alpha}} + f_{i}
,
\label{eq:eqforx}
\end{equation}
where a sum over the repeated Greek indices is assumed. 
$S$ is the entropy of the entire system.  
$K_{ij} \equiv M_{ij}+L_{ij}$ is a kinetic coefficient
which generally depends on ${\bf x}$.
$M_{i\alpha}\partial S/\partial x_{\alpha}$ represents the reversible term
where $M_{ij}=-M_{ji}$ is an antisymmetric matrix.
$L_{i\alpha}\partial S/\partial x_{\alpha}$ represents the irreversible
term, where $L_{ij}$ is the (${\bf x}$-dependent) Onsager coefficient. 
$f_{i}({\bf x},t)$ is a random noise which satisfies 
\begin{equation}
\left\langle f_{i}({\bf x},t)f_{j}({\bf x}',t') \right\rangle_{{\bf x}(t)={\bf x}}
= 2k_{\mbox{\scriptsize B}}  L_{ij}({\bf x})\delta(t-t'), 
\label{eq:2FDT}
\end{equation}
where $\left\langle \cdots \right\rangle_{{\bf x}(t)={\bf x}}$ denotes the 
conditional average in which the ensemble average is taken with 
a fixed value of ${\bf x}(t)={\bf x}$ at time $t$.
The fact that the Onsager coefficient is a function of ${\bf x}$ 
means that the random noise is also the function of ${\bf x}$,
{\it i.e.}, it is multiplicative\cite{vankampen1981}. 
Eq.(\ref{eq:eqforx}) is the general expression for 
dynamical systems 
whose stationary distribution function in the absence of  nonequilibrium
constraints is given by the equilibrium ensemble. 
Examples include  the (fluctuating) Navier-Stokes
equation\cite{landau1959,Miyazaki1996} and 
the nonlinear diffusion equation (eq.(\ref{eq:dean})). 
The response function $\chi_{ij}(t)$ 
is defined as the response to a time-dependent external force ${\bf
F}(t)$  
by
\begin{equation}
\langle \Delta x_{i}(t)\rangle_{F}
= \int_{-\infty}^{t}\!\!\!\!{\mbox{d}} t'~\chi_{i\alpha}(t-t')F_{\alpha}(t'),
\label{eq:responsedef}
\end{equation}
where 
$\langle\Delta x_{i}(t)\rangle_{F}$ is the deviation of 
$\langle x_{i}(t)\rangle$ 
from its equilibrium value due to the external force.
The FDT asserts that $\chi_{ij}(t)$ is related to 
the correlation function in the absence of ${\bf F}(t)$, 
$C_{ij}(t) = \left\langle x_{i}(t)x_{j}(0)\right\rangle$, 
by
\begin{equation}
\chi_{ij}(t)= -\frac{1}{k_{\mbox{\scriptsize B}} T}\diff{C_{ij}(t)}{t}
\hspace*{1.0cm}{\mbox{for~}} t \geq 0.
\label{eq:FDT}
\end{equation}
The FDT is proved easily using linear response
theory\cite{graham1973}. 
The FDT is one of the strongest and the most robust
statements of equilibrium statistical physics 
and it holds for any dynamical
processes, classical or quantum, linear or nonlinear, 
as long as the system is stationary, satisfies the condition of  
detailed balance, and the perturbation is small enough.   

The MSR formalism allows us for the use of a systematic loop expansion  
for the solution of 
the nonlinear Langevin equation in terms of the moments. 
MCT is regarded as the lowest order 
self-consistent approximation with no vertex correction in the MSR 
formalism.   
Deker {\it et al. }\cite{deker1975} have proven that the FDT holds 
at each order of the loop expansion for three special classes of 
dynamical processes: 
``Class A'' where, in eq.(\ref{eq:eqforx}), 
$M_{ij}({\bf x})=0$,  $L_{ij}$ is a constant (independent of ${\bf x}$) 
and thus the noise is additive. 
The nonlinearity of the Langevin equation originates from the entropy
$S$. 
``Class B'' where $L_{ij}$ is constant and  the entropy is a quadratic
function of ${\bf x}$.  
The reversible matrix $M_{ij}({\bf x})$ depends on ${\bf x}$ which is the origin
of the nonlinearity. 
The equations discussed by Kawasaki to describe dynamical critical
phenomena\cite{kawasaki1970} belong to this class. 
``Class C'' involves Hamiltonian systems which do not have an
irreversible part. 

The problems is that, as discussed in the Introduction, even 
eq.(\ref{eq:dean}), which describes the dynamics of
supercooled fluids, does not 
belong to any of the classes listed above.  
The nonlinear term in eq.(\ref{eq:dean}) 
originates from the combination of the non-quadratic term
of the entropy and the variable dependence of the Onsager coefficient.
Extensions of the MSR formalism to more general cases have 
been discussed in 
Refs.\cite{kawasaki1997e,deker1979,phythian1977,jensen1981}
but derivations of MCT equations have not
been given so far. 
Since Deker's classifications do not cover these dynamical
processes, 
it is convenient to re-categorize 
the nonlinear stochastic processes in a slightly
more general way than
Classes A--C of Deker:

\vspace*{0.1cm}

\noindent
(Class I):~
The nonlinearity comes solely from the entropy. 
$K_{ij}$ is independent of ${\bf x}$. 
The noise is additive.
The mean-field model of $p$-spin spin glasses also belongs to this
type\cite{bouchaud1996}. 
Again, it is trivial to show that the FDT holds at each order of the   
loop expansion. 

\vspace*{0.1cm}

\noindent
(Class II):~
The entropy is a quadratic function of ${\bf x}$ but $K_{ij}({\bf x})$
is dependent on ${\bf x}$. 
$L_{ij}({\bf x})$ can be also a function of ${\bf x}$ and, therefore,
the noise can be multiplicative. 

\vspace*{0.1cm}

\noindent
(Class III):~
	    The entropy is an arbitrary function of ${\bf x}$ and 
	    $K_{ij}({\bf x})$ is dependent on ${\bf x}$. 
Real fluids including the one described by eq.(\ref{eq:dean}) belong
to this type.

\vspace*{0.1cm}

\noindent
In this letter, we
shall focus here on Class II.  
Here we restrict ourselves to the simplest situation where 
$K_{ij}({\bf x})$ is a linear function of $\delta{\bf x}$: 
\begin{equation}
K_{ij}({\bf x})= K_{ij}^{(0)} + K_{ij,\alpha}^{(1)}\delta x_{\alpha},  
\label{eq:K}
\end{equation}
because this is the most important case in the context of the glass
transition. 
We shall show, for this class of Langevin equation, that MCT is
consistent  with the FDT even with the presence of multiplicative noise. 
We believe that this conclusion is valid for arbitrary function  of
$K_{ij}({\bf x})$ of Class II.  
We will end this work with some comments on the more interesting Class
III case which is problematic from the standpoint of the FDT within the
MCT approximation.

\section{MSR formalism for processes with multiplicative noise}

In this section, we shall develop the MSR method for the Class II
case.  
In Class II, the entropy in eq.(\ref{eq:eqforx}) is
given by a quadratic form 
\begin{equation}
S = S_{0} 
    + \frac{1}{2}\Omega_{\alpha\beta}\delta x_{\alpha}\delta x_{\beta},
\label{eq:GaussS}
\end{equation}
where $\delta{\bf x} ={\bf x}-\langle{\bf x}\rangle$ and 
$\Omega_{ij}^{-1}=-k_{\mbox{\scriptsize B}}C_{ij}^{-1}(t=0)$ 
is the inverse of the equal time correlation function. 
Such an approximation is, perhaps, not as crude as it may superficially
appear. 
Indeed, it has been shown via direct simulation that both
simple\cite{Crooks1997} and complex liquids\cite{Hummer1996} have  
Gaussian density fluctuations over a wide range of length scales. 
Eq.(\ref{eq:GaussS}), within the canonical ensemble, is a precise
statement of this approximation.  
The Fokker-Planck equation for the probability density function 
$P({\bf x},t)$ 
equivalent with eq.(\ref{eq:eqforx}) is written as 
\begin{equation}
\fl\hspace*{1.0cm}
\pdif{P({\bf x}, t)}{t}
= -\pdif{~}{x_{\alpha}}\left\{
    K_{\alpha\beta}({\bf x})\pdif{S}{x_{\beta}} 
-k_{\mbox{\scriptsize B}} L_{\alpha\beta}({\bf x})\pdif{~}{x_{\beta}}
\right\} P({\bf x}, t)
\equiv {\cal T} P({\bf x}, t).
\label{eq:FP}
\end{equation}
For the detailed balance condition to be satisfied, 
$M_{ij}({\bf x})$ must satisfy the following condition 
(potential condition\cite{graham1973}):  
$\partial{M_{i\alpha}({\bf x})}/\partial{x_{\alpha}}=0$.
We shall also assume the similar incompressible condition for 
$L_{ij}({\bf x})$: 
$\partial{L_{i\alpha}({\bf x})}/\partial{x_{\alpha}}=0$.
These two conditions are satisfied for most 
hydrodynamic equations including eq.(\ref{eq:dean}) (see
eq.(\ref{eq:Lij})).  
The latter condition is especially useful because it enables us to avoid
distinguishing between the  
Ito and Stratonovich interpretations which are associated with the
multiplicative noise\cite{vankampen1981}.   
%
Using eqs.(\ref{eq:K}) and (\ref{eq:GaussS}), eq.(\ref{eq:eqforx})
can be rewritten as  
\begin{equation}
\eqalign{
\dot{x}_{i} 
&
=
\mu_{i\alpha}x_{\alpha}+\frac{1}{2}{\cal V}_{i\alpha\beta}x_{\alpha}x_{\beta}
+f_{i}.
} 
\label{eq:expansion}
\end{equation}
Hereafter we shall omit the ``$\delta$'' in front of ${\bf x}$. 
In eq.(\ref{eq:expansion}),
$\mu_{ij}={K}^{(0)}_{i\alpha}\Omega_{\alpha j}$ 
is a bare transport coefficient and 
${\cal V}_{ijk}={\cal V}_{ikj}$ is the symmetrized vertex defined by 
\begin{equation}
\fl\hspace*{1.0cm}
{\cal V}_{ijk}
=
  M^{(1)}_{i\alpha,j}\Omega_{\alpha k}
+ M^{(1)}_{i\alpha,k}\Omega_{\alpha j}
+ L^{(1)}_{i\alpha,j}\Omega_{\alpha k}
+ L^{(1)}_{i\alpha,k}\Omega_{\alpha j}
\equiv
{\cal M}_{ijk} + {\cal L}_{ijk},
\label{eq:Vijk}
\end{equation}
where ${\cal M}_{ijk}$ and ${\cal L}_{ijk}$ are 
reversible and irreversible contributions of ${\cal V}_{ijk}$,
respectively.  
${\cal M}_{ijk}$ satisfies the cyclic condition given by 
\begin{equation}
 \Omega_{i\alpha}{\cal M}_{\alpha jk}
+\Omega_{j\alpha}{\cal M}_{\alpha ki}
+\Omega_{k\alpha}{\cal M}_{\alpha ij}=0.
\label{eq:cyclic}
\end{equation}
This is proved  using the condition that the reversible part does not
contribute to the entropy production. 

Following the standard MSR procedure\cite{martin1973}, we shall
introduce a spinor 
${\bf z} \equiv ({\bf x}, {\hat{\bf{x}}})$, where 
$\hat{x}_{i} \equiv -\partial/\partial x_{i}$. 
It is convenient to define the generating function for ${\bf z}$ by
\begin{equation}
W[{\mbox{\boldmath$\xi$}}]
\equiv 
\ln
\left\langle \exp_{+}
\left[ \int\!\!{\mbox{d}}t~ 
{\mbox{\boldmath$\xi$}}\cdot{\bf z} \right]
\right\rangle
,
\end{equation}
where 
${\mbox{\boldmath$\xi$}}
=({\mbox{\boldmath$\eta$}}, \hat{\mbox{\boldmath$\eta$}})$ 
is the auxiliary field conjugate to ${\bf z}$ 
which will be eventually set to zero.
``$\exp_{+}$'' implies the time ordering which aligns the quantities
with larger $t$ on the left.
We define the cumulant function 
${\cal G}_{\xi}(1,\cdots,n)
=\left\langle \!\left\langle z(1)\cdots z(n) \right\rangle\!\right\rangle$
by 
\begin{equation}
{\cal G}_{\xi}(1,\cdots,n)
=
\frac{\delta^{n}W[{\mbox{\boldmath$\xi$}}]}{\delta\xi(1)\cdots \delta\xi(n)}.
\label{eq:Gn}
\end{equation}
The index number $1 = ( i, t, \pm)$, etc...
represents the index for field variables $i$,
time $t$, and the index of the spinor 
defined by  ${\bf z}(+)={\bf x}$ and ${\bf z}(-)={\hat{\bf{x}}}$, respectively.
Let us construct the equation of motion for the first cumulant, 
$\left\langle \!\left\langle z(1) \right\rangle\!\right\rangle$. 
Substituting eq.(\ref{eq:expansion}) into eq.(\ref{eq:Gn}) for $n=1$, we
obtain the Schwinger equation:
\begin{equation}
\diff{\left\langle \!\left\langle z(1) \right\rangle\!\right\rangle}{t}
=
- \left\langle \!\left\langle \left[ {\cal T}, z(1) \right] \right\rangle\!\right\rangle
+ i\sigma(1,\underline{1})\xi(\underline{1}),
\label{eq:schwinger}
\end{equation}
where a sum over the repeated underlined indices is assumed. 
$[\cdots]$ is the commutator, 
${\cal T}$ is the Fokker-Planck operator defined by 
eq.(\ref{eq:FP}) represented in terms of 
$({\bf x},\hat{\bf{x}})$, 
and 
\begin{equation}
\eqalign{
i\sigma(1,2)
&
= \left( 
\begin{array}{cc}
{\bf 0}  & {\bf -1} \\
{\bf 1}  & {\bf 0}  \\
\end{array} 
\right)\delta(t_{1}-t_{2})
.  
}
\end{equation}
The explicit expression of eq.(\ref{eq:schwinger}) is given by 
\begin{equation}
\fl\hspace*{0.5cm}
\eqalign{
\diff{\left\langle \!\left\langle x_{i} \right\rangle\!\right\rangle}{t}
=
&
 {\hat{\eta}}_{i} + \mu_{i\alpha}\left\langle \!\left\langle x_{\alpha} \right\rangle\!\right\rangle
 + \frac{1}{2}{\cal V}_{i\alpha\beta}
 \left\{ \left\langle \!\left\langle x_{\alpha}x_{\beta}\right\rangle\!\right\rangle
        +\left\langle \!\left\langle x_{\alpha}\right\rangle\!\right\rangle\left\langle \!\left\langle x_{\beta}\right\rangle\!\right\rangle
 \right\}
\\
&
+2k_{\mbox{\scriptsize B}}\left[
           L^{(0)}_{i\alpha}\left\langle \!\left\langle \hat{x}_{\alpha}\right\rangle\!\right\rangle
          +L^{(1)}_{i\alpha,\beta}
           \left\{ \left\langle \!\left\langle \hat{x}_{\alpha}x_{\beta}\right\rangle\!\right\rangle
                  +\left\langle \!\left\langle \hat{x}_{\alpha}\right\rangle\!\right\rangle\left\langle \!\left\langle x_{\beta}\right\rangle\!\right\rangle
          \right\} 
     \right]
\\
\diff{\left\langle \!\left\langle \hat{x}_{i} \right\rangle\!\right\rangle}{t}
=
&
-\eta_{i}
-{}^{\mbox{\scriptsize t}}\!\mu_{i\alpha}
\left\langle \!\left\langle 
\hat{x}_{\alpha} 
\right\rangle\!\right\rangle
-{\cal V}_{\alpha\beta i}
\left\{ \left\langle \!\left\langle \hat{x}_{\alpha}x_{\beta} \right\rangle\!\right\rangle
       +\left\langle \!\left\langle \hat{x}_{\alpha}\right\rangle\!\right\rangle\left\langle \!\left\langle x_{\beta}\right\rangle\!\right\rangle
\right\}
+{\cal V}_{\alpha\alpha i}
\\
&
-k_{\mbox{\scriptsize B}} L^{(1)}_{\alpha\beta,i}
\left\{ 
\left\langle \!\left\langle \hat{x}_{\alpha}\hat{x}_{\beta} 
\right\rangle\!\right\rangle
+
\left\langle \!\left\langle \hat{x}_{\alpha}\right\rangle\!\right\rangle
\left\langle \!\left\langle \hat{x}_{\beta} \right\rangle\!\right\rangle
\right\},
} 
\label{eq:schwinger2}
\end{equation}
where ${}^{\mbox{\scriptsize t}}\!\mu_{ij}$ is the transverse of
$\mu_{ij}$. 
The last terms in these equations are due to the multiplicative noise.
Eq.(\ref{eq:schwinger2}) is written in short as
\begin{equation}
\fl\hspace*{0.5cm}
{\cal G}_{0}^{-1}(1,\underline{1})
\left\langle \!\left\langle z(\underline{1}) \right\rangle\!\right\rangle
=
\xi(1)+ C(1)
+\frac{1}{2}\gamma_{3}(1,\underline{1},\underline{2})
\left\{
 \left\langle \!\left\langle z(\underline{1})z(\underline{2}) \right\rangle\!\right\rangle
+\left\langle \!\left\langle z(\underline{1})\right\rangle\!\right\rangle
 \left\langle \!\left\langle z(\underline{2})\right\rangle\!\right\rangle
\right\}
,
\label{eq:schwinger3}
\end{equation}
where 
$C(1)\equiv ({\cal V}_{\alpha\alpha i},0)$ is a constant which does not
contribute to the following arguments and
${\cal G}_{0}(1,2)$ is the bare propagator whose inverse is written as
\begin{equation}
{\cal G}_{0}^{-1}(1,2)
= i\sigma(1,2) \diff{~}{t_2} + \gamma_{2}(1,2)
\label{eq:G0}
\end{equation}
with the symmetric matrix $\gamma_{2}(1,2)$ defined by 
\begin{equation}
\eqalign{
\gamma_{2}(1,2)
&
= 
\left( 
\begin{array}{cc}
0     & -{}^{\mbox{\scriptsize t}}\!{\mbox{\boldmath$\mu$}} \\
-{\mbox{\boldmath$\mu$}} & -2k_{\mbox{\scriptsize B}}{\bf L}^{(0)} \\
\end{array}
\right)\delta(t_1-t_2).
}
\label{eq:coeff}
\end{equation}
The non-zero components of $\gamma_{3}(1,2,3)$ are
\begin{equation}
\fl\hspace*{1.0cm}
\left\{ 
\eqalign{
&
\gamma_{3}(i_1,t_1,-; i_2,t_2,+;i_3,t_3,+)
={\cal V}_{i_{1}i_{2}i_{3}}\delta(t_1-t_2)\delta(t_1-t_3)
\\
&
\gamma_{3}(i_1,t_1,-; i_2,t_2,-;i_3,t_3,+)
= 2k_{\mbox{\scriptsize B}} L^{(1)}_{i_{1}i_{2}, i_{3}}
\delta(t_1-t_2)\delta(t_1-t_3)
} 
\right.
\label{eq:g3}
\end{equation}
and its permutation of the indices $(1,2,3)$. 
Note that  $\gamma_{3}(1,2,3)$ is a fully symmetric tensor.
The second moment 
${\cal G}(1,2)={\cal G}_{\xi=0}(1,2)=\langle z(1)z(2) \rangle$
is given by taking the derivative of eq.(\ref{eq:schwinger3}) 
with respect to 
$\left\langle \!\left\langle z(2) \right\rangle\!\right\rangle$ 
using eq.(\ref{eq:Gn}) and then turning off 
${\mbox{\boldmath$\xi$}}=0$: 
\begin{equation}
{\cal G}^{-1}(1,2)={\cal G}_{0}^{-1}(1,2)-\Sigma(1,2),
\label{eq:MCT}
\end{equation}
where $\Sigma(1,2)$ is the self energy. 
By neglecting the vertex correction, we obtain the MCT expression for
the self-energy: 
\begin{equation}
\Sigma(1,2)
\simeq
\frac{1}{2}\gamma_{3}(1,\underline{1},\underline{2})
{\cal G}(\underline{1},\underline{3}){\cal G}(\underline{2},\underline{4})
\gamma_{3}(2,\underline{3},\underline{4}).
\label{eq:Sigma}
\end{equation}
We can write the components of these matrices as 
\begin{equation}
{\cal G}(1,2)
\equiv 
\left( 
\begin{array}{ll}
{\bf C}  & {\bf G}  \\
{\bf G}^{\dagger}  & {\bf 0}  \\
\end{array}
\right)
\hspace*{0.5cm}
\mbox{and}
\hspace*{0.5cm}
\Sigma(1,2)
\equiv 
\left( 
\begin{array}{cc}
{\bf 0}  & {\bf E}^{\dagger} \\
{\bf E}       & {\bf D} \\
\end{array}
\right)
,
\end{equation}
where ``$\dagger$'' represents Hermitian conjugate defined by 
$A^{\dagger}_{ij}(t-t')=A_{ji}^{\ast}(t'-t)$.
From the structure of eq.(\ref{eq:MCT}), it is straightforward to
show that  
${\cal G}(-,-)=\Sigma(+,+)=0$.
$C_{ij}(t-t')=\langle x_{i}(t)x_{j}(t')\rangle$ 
is the correlation function and 
${G}_{ij}(t-t')=\langle x_{i}(t)\hat{x}_{j}(t')\rangle$ 
is the propagator which describes the response of the system to the
random noise.  
For $t> 0$, the equations for ${\bf C}(t)$ and ${\bf G}(t)$ 
can be written explicitly using
eqs.(\ref{eq:G0})-(\ref{eq:MCT}) as
\begin{eqnarray}
&
\fl\hspace*{0.5cm}
\diff{C_{ij}(t)}{t}
= 
\mu_{i\alpha}C_{\alpha j}(t)
+
\int_{-\infty}^{t}\!\!\!\!\!\!{\mbox{d}} t_1~
{E}_{i\alpha}(t-t_1)C_{\alpha j}(t_1) 
+ 
\int_{-\infty}^{0}\!\!\!\!\!\!{\mbox{d}} t_1~
D_{i\alpha}(t-t_1){G}^{\dagger}_{\alpha j}(t_1)
\label{eq:MCTC}
\\
&
\fl\hspace*{0.5cm}
\diff{{G}_{ij}(t)}{t}
= 
\mu_{i\alpha}{G}_{\alpha j}(t)
+\int_{0}^{t}\!\!{\mbox{d}} t_1~{E}_{i\alpha}(t-t_1){G}_{\alpha j} (t_1)
\label{eq:MCTG}
\end{eqnarray}
with the self energies given by
\begin{equation}
\fl\hspace*{1.0cm}
\left\{
\eqalign{
{E}_{ij}(t)= 
&
{\cal V}_{i\alpha\beta}{G}_{\alpha\lambda}(t)C_{\beta\mu}(t){\cal V}_{\lambda\mu j}
+
k_{\mbox{\scriptsize B}} {\cal V}_{i\alpha\beta}{G}_{\alpha\lambda}(t){G}_{\beta\mu}(t)
L^{(1)}_{\lambda\mu, j}
\\
D_{ij}(t)= 
&
\frac{1}{2}{\cal V}_{i\alpha\beta}
C_{\alpha\lambda}(t)C_{\beta\mu}(t)
{\cal V}_{j \lambda\mu}
+2k_{\mbox{\scriptsize B}} {\cal V}_{i\alpha\beta}G_{\alpha\lambda}(t)C_{\beta\mu}(t)
L^{(1)}_{j \lambda, \mu},
}\right.
\label{eq:E-D}
\end{equation}
where use has been made of the causality condition:
${\bf G}(t)=0$ for $t < 0$. 
The terms containing $L_{ij,k}^{(1)}$ in eq.(\ref{eq:E-D}) originate
from the multiplicative noise. 
Similar terms were derived by  Kawasaki {\it et al.}\cite{kawasaki1997e}
but they were disregarded and their importance was not addressed. 
{\it Note that the propagator ${\bf G}(t)$ represents the response 
to the noise but it is not the response to the external force
defined by eq.(\ref{eq:responsedef}).} 
The response function is obtained by evaluating the 
linear response of the average $\langle x_{i}(t) \rangle$ to the external
force ${\bf F}(t)$. 
The term associated with the external force is introduced naturally by
replacing the entropy 
with the one associated with the work done by the force as
\begin{equation}
S_{F} = S +\frac{{\bf x}\cdot{\bf F}}{T}
.
\end{equation}
Inserting this expression into the entropy term in eq.(\ref{eq:FP}) and
taking the leading order of the formal solution, it is straightforward
to derive the expression for the response function. The result is
\begin{equation}
\fl\hspace*{1.0cm}
\eqalign{
\chi_{ij}(t-t')
&
= 
\frac{1}{T}\langle x_{i}(t)\hat{x}_{\alpha}(t'){K}_{\alpha j}({\bf x}(t'))
           \rangle
\\
&
=
\frac{1}{T}\langle x_{i}(t)\hat{x}_{\alpha}(t')\rangle{K}_{\alpha j}^{(0)}
+\frac{1}{T}
\langle x_{i}(t)\hat{x}_{\alpha}(t')x_{\beta}(t')\rangle{K}_{\alpha j,\beta}^{(1)}.
}
\end{equation}
The three point correlation function 
$\langle x_{i}(t)\hat{x}_{\alpha}(t')x_{\beta}(t') \rangle$
in this expression is calculated in 
the same spirit as derivation
of eq.(\ref{eq:Sigma}). 
Up to the one-loop level, neglecting the vertex correction, it is
written as 
$\langle z(1)z(2)z(3) \rangle
\simeq {\cal G}(1,\underline{1}){\cal G}(2,\underline{2})
{\cal G}(3,\underline{3})\gamma_{3}(\underline{1},\underline{2},\underline{3})$.
Substituting eq.(\ref{eq:g3}) to this, one obtains the MCT expression
of the response function: 
\begin{equation}
\fl\hspace*{1.0cm}
\eqalign{
&
\chi_{ij}(t)
=
\frac{1}{T}{G}_{i\alpha}(t){K}_{\alpha j}^{(0)}
+ 
\frac{1}{T}
\int_{0}^{t}\!\!{\mbox{d}} t_{1}~
{G}_{i\alpha}(t-t_{1})
{\cal V}_{\alpha\beta\gamma}
{G}_{\beta\lambda}(t_{1})C_{\gamma\mu}(t_{1})K_{\lambda j,\mu}^{(1)} 
.
} 
\label{eq:chi}
\end{equation}
In previous works, the propagator $T^{-1}{\bf G}(t)\cdot{\bf K}^{(0)}$ 
has been called the response function\cite{deker1975}. 
But as discussed above, it is not identical to the full response
function in general. 
They become identical only if the kinetic coefficient $K_{ij}$ is a
constant, {\it i.e.}, for Class I.

\section{FDT and MCT for Class II}

In this section, we prove that the correlation and response functions
given by eqs.(\ref{eq:MCTC}), (\ref{eq:MCTG}),  and (\ref{eq:chi})
satisfy the FDT, 
eq.(\ref{eq:FDT}). 
Deker {\it et al.} have shown that, for the MCT equation of
Class B, there is a simple relation between the correlation function
and the propagator if $L_{ij}$ is a constant or
$L_{ij,k}^{(1)}=0$\cite{deker1975}:   
\begin{equation}
{\bf G}(t) = \theta(t){\bf C}(t)\cdot{\bf C}^{-1}(0),
\label{eq:G2C}
\end{equation}
where $\theta(t)$ is the Heaviside function. 
We prove that this is also true when $L_{ij,k}^{(1)}\neq 0$ or 
the noise is multiplicative as follows:
Taking the time derivative of both side of eq.(\ref{eq:G2C})
and substituting the equation for the correlation function,
eq.(\ref{eq:MCTC}), we have
\begin{equation}
\diff{{\bf G}}{t}
= {\bf 1}+ {\mbox{\boldmath$\mu$}}\cdot{\bf G}  + \theta(t)
 \left\{ {\bf E}\otimes{\bf C}  + {\bf D}\otimes{\bf G}^{\dagger} 
 \right\}\cdot{\bf C}^{-1}(0)
,
\label{eq:G}
\end{equation}
where 
${\bf A}\otimes{\bf B} \equiv 
\int_{-\infty}^{\infty}\mbox{d}t_1~A_{i\alpha}(t-t_1)B_{\alpha j}(t_1)$. 
The terms containing ${\cal M}_{ijk}$ in 
${\bf E}\otimes{\bf C}\cdot{\bf C}^{-1}(0)$ 
can be rearranged using eq.(\ref{eq:G2C}) and the cyclic condition,
eq.(\ref{eq:cyclic}).  
For example, if $t_1 \leq 0$, 
\begin{equation}
\fl\hspace*{1.0cm}
{\cal V}_{i\alpha\beta}{G}_{\alpha\lambda}(\tau)C_{\beta\mu}(\tau)
{\cal M}_{\lambda\mu\nu}C_{\nu j}(t_{1})
= 
-\frac{1}{2}
{\cal V}_{i\alpha\beta}C_{\alpha\lambda}(\tau)C_{\beta\mu}(\tau)
{\cal M}_{\nu\lambda\mu}{G}^{\dagger}_{\nu j}(t_{1}),
\label{eq:SEM}
\end{equation}
where $\tau=t-t_1$. 
Here we have used the fact that 
$C_{ij}(0)=-k_{\mbox{\scriptsize B}}\Omega_{ij}^{-1}$. 
On the other hand, the terms containing ${\cal L}_{ijk}$ 
in ${\bf E}\otimes{\bf C}\cdot{\bf C}^{-1}(0)$ 
are rearranged as  
\begin{equation}
\fl\hspace*{1.0cm}
\eqalign{
&
\left\{
{G}_{\alpha\lambda}(\tau)C_{\beta\mu}(\tau){\cal L}_{\lambda\mu k}
+
k_{\mbox{\scriptsize B}}{G}_{\alpha\lambda}(\tau){G}_{\beta\mu}(\tau)
L^{(1)}_{\lambda\mu,  k}
\right\}C_{kj}(t_1)
\\
&
=
\left\{
-2k_{\mbox{\scriptsize B}} 
{G}_{\alpha\lambda}(\tau)C_{\beta\mu}(\tau)L^{(1)}_{k\lambda,\mu}
-\frac{1}{2}C_{\alpha\lambda}(\tau)C_{\beta\mu}(\tau){\cal L}_{k\lambda\mu} 
\right\}
{G}^{\dagger}_{kj}(t_1).
} 
\label{eq:SEL}
\end{equation}
Eq.(\ref{eq:SEM}) combined with eq.(\ref{eq:SEL})
cancels with 
${\bf D}\otimes{\bf G}^{\dagger}\cdot{\bf C}^{-1}(0)$ 
of eq.(\ref{eq:G}).
Likewise, for $t_1\geq 0$, 
${\bf E}\otimes{\bf C}\cdot{\bf C}^{-1}(0)$ 
can be rewritten as 
${\bf E}\otimes{\bf G}^{\dagger}$.
Therefore, eq.(\ref{eq:G}) becomes equivalent to the equation
for ${\bf G}$, eq.(\ref{eq:MCTG}). 
This is the end of the proof. 

Now let us prove the FDT. 
By taking the derivative of eq.(\ref{eq:G2C})
with respect to time and using equation for ${\bf G}(t)$,
eq.(\ref{eq:MCTG}), we have  
\begin{equation}
\diff{{\bf C}(t)}{t}
= -k_{\mbox{\scriptsize B}} {\bf G}(t)\cdot{\bf K}^{(0)}
+\int_{0}^{t}\!\!{\mbox{d}} t_1~{\bf G}(t-t_1)\cdot{\bf E}(t_1)\cdot{\bf C}(0). 
\label{eq:dCdt}
\end{equation}
In this expression, ${\bf E}(t)\cdot{\bf C}(0)$ can be rewritten
using eq.(\ref{eq:G2C}) and antisymmetric property of $M_{ij,k}^{(1)}$ 
as
\begin{equation}
\eqalign{
\left\{ {\bf E}(t)\cdot{\bf C}(0)\right\}_{ij}
= 
&
-k_{\mbox{\scriptsize B}} {\cal V}_{i\alpha\beta}{G}_{\alpha\lambda}(t)C_{\beta\mu}(t)
K^{(1)}_{\lambda j, \mu}
.
} 
\label{eq:EC}
\end{equation}
Therefore, the right hand side of eq.(\ref{eq:dCdt}) becomes identical
to $-k_{\mbox{\scriptsize B}} T\chi_{ij}(t)$ given by eq.(\ref{eq:chi}). 
Thus we arrive at eq.(\ref{eq:FDT}) and the FDT is proved. 

Finally let us derive the closed equation for $C_{ij}(t)$. 
${\bf D}\otimes{\bf G}^{\dagger}$ again cancels with 
${\bf E}\otimes{\bf C}$ for $t_1 \leq 0$. 
For $t_1 \geq 0$,  
${E}_{ij}(t)$ is rewritten as 
\begin{equation}
{E}_{ij}(t)
= 
-\frac{1}{2}{\cal V}_{i\alpha\beta}C_{\alpha\lambda}(t)C_{\beta\mu}(t)
\left( {\cal V}-2{\cal L} \right)_{\nu \lambda\mu}C^{-1}_{\nu j}(0)
\label{eq:E-mct}
\end{equation}
and we arrive at the MCT equation for $C_{ij}(t)$;
\begin{equation}
\diff{C_{ij}(t)}{t}
= \mu_{i\alpha}C_{\alpha j}(t) +
\int_{0}^{t}\!\!{\mbox{d}} t_{1}~{E}_{i\alpha}(t-t_1)C_{\alpha j}(t_1).
\label{eq:mct}
\end{equation}
{\it It is important to realize 
that $-2{\cal L}$ in the vertex in eq.(\ref{eq:E-mct}) 
is due to the multiplicative noise and the presence of it
is essential. }
For example, for the pure dissipative case ($M_{ij}({\bf x})=0$), 
neglect of the multiplicative noise leads to the wrong sign 
in front of the integral term
(and thus leads to pathological behavior). 
This term is neglected in Ref.\cite{kawasaki1997e}. 

%

\section{FDT and MCT for Class III}

In this section, we shall consider the category of problems we call
Class III and elucidate  
generic reason why the MCT approximation for the Class III dynamics is
inconsistent with the FDT.  
For complete discussion of the technical aspects involved would require
a longer discussion than we provide here. 
Our main point is simply to sketch the difficulties that arise in
attempting to formulate a simple MCT (namely a self-consistent one-loop 
theory for both the propagation and the response function) that
satisfies the FDT. 
The important conclusion is that the standard idealized MCT of G\"{o}tze
and coworkers\cite{gotze1992,gotze1999} can not be consistently derived
via field-theoretic techniques, at least via the usual one-loop
approximations applied to eq.(\ref{eq:dean}). 
This is discussed both in this section and in Section 6.

For problems of Class III, the entropy is not a quadratic function but 
has higher order terms;
\begin{equation}
S = S_{0} 
    + \frac{1}{2}\Omega_{\alpha\beta}\delta x_{\alpha}\delta x_{\beta}
    + \frac{1}{3!}\Lambda_{\alpha\beta\gamma}
    \delta x_{\alpha}\delta x_{\beta}\delta x_{\gamma} 
    + \cdots 
.
\label{eq:higherS}
\end{equation}
We again assume that the kinetic coefficient $K_{ij}({\bf x})$ is a
linear function of ${\bf x}$. 
Up to the quadratic order in ${\bf x}$, the nonlinear Langevin equation
for ${\bf x}$ is given by eq.(\ref{eq:expansion}) but the vertex,
eq.(\ref{eq:Vijk}), is now modified as
\begin{equation}
\fl\hspace*{1.0cm}
\eqalign{
{\cal V}_{ijk}
&
=
  K^{(0)}_{i\alpha}\Lambda_{\alpha jk}
+ M^{(1)}_{i\alpha,j}\Omega_{\alpha k}
+ M^{(1)}_{i\alpha,k}\Omega_{\alpha j}
+ L^{(1)}_{i\alpha,j}\Omega_{\alpha k}
+ L^{(1)}_{i\alpha,k}\Omega_{\alpha j}
\\
&
\equiv
{\cal V}^{(I)}_{ijk}+{\cal V}^{(II)}_{ijk}, 
}
\label{eq:Vijk3}
\end{equation}
where ${\cal V}^{(I)}_{ijk}\equiv K^{(0)}_{i\alpha}\Lambda_{\alpha jk}$
is the vertex that originates from the nonlinearity of the entropy 
whereas
${\cal V}^{(II)}_{ijk}$, which is the same as eq.(\ref{eq:Vijk}),
originates from the ${\bf x}$-dependence of the kinetic coefficient
$K_{ij}({\bf x})$. 
The one-loop equations for $C_{ij}(t)$ and $G_{ij}(t)$,
eqs.(\ref{eq:MCTC}) and (\ref{eq:MCTG}), remain the same except that the
vertex is given by 
eq.(\ref{eq:Vijk3}) instead of eq.(\ref{eq:Vijk}). 
First, we shall show that there is no simple relation such as eq.(\ref{eq:G2C})
which relates the correlation function to the propagator. 
The starting point is the formal solution of eq.(\ref{eq:MCTC}) for the
correlation function; 
\begin{equation}
{\bf C} = 
{\bf G}\otimes 
\left( k_{\mbox{\scriptsize B}}{\bf K}^{(0)} 
      +k_{\mbox{\scriptsize B}}{\bf K}^{(0)\dagger} + {\bf D} \right)
\otimes {\bf G}^{\dagger}. 
\label{eq:Cformal}
\end{equation} 
This is derived by eliminating ${\bf E}(t)$ from the formal solution 
of ${\bf C}(t)$ by substituting the formal solution for ${\bf G}(t)$. 
Substituting the equation for ${\bf G}$, eq.(\ref{eq:MCTG}), 
eq.(\ref{eq:Cformal}) can be rewritten as 
\begin{equation}
{\bf C} = 
-k_{\mbox{\scriptsize B}}{\bf G}\cdot{\mbox{\boldmath$\Omega$}}^{-1}
+ 
{\bf G}\otimes{\bf f}\otimes{\bf G}^{\dagger}   
\label{eq:CformalIII}
\end{equation} 
for $t \geq 0$. 
In this expression, 
${\bf f}(t) \equiv 
-k_{\mbox{\scriptsize B}}{\bf E}(t)\cdot{\mbox{\boldmath$\Omega$}}^{-1}
-k_{\mbox{\scriptsize B}}{\mbox{\boldmath$\Omega$}}^{-1}\cdot{\bf E}(t)
+ {\bf D}(t)$. 
Following the similar steps as eqs.(\ref{eq:SEL}--\ref{eq:EC}), 
${\bf f}(t)$ can be reduced as 
\begin{equation}
\fl\hspace*{1.0cm}
\eqalign{
f_{ij}(t)
= 
&
-{\cal V}_{i\alpha\beta}{G}_{\alpha\lambda}(t)C_{\beta\mu}(t)
{\cal V}^{(I)}_{\lambda\mu\nu}\Omega_{\nu j}^{-1}
+
\frac{1}{2}
{\cal V}_{i\alpha\beta}{C}_{\alpha\lambda}(t)C_{\beta\mu}(t)
{\cal V}^{(I)}_{j\lambda\mu}
\\
&
+ \mbox{(Higher order loops)},  
}
\end{equation}
where only the first two terms (one loop) are explicitly shown. 
The other terms consist of the higher order loops. 
These loops appear to contain at least one multiple of 
${\cal V}^{(I)}{\cal V}^{(II)}$. 
This means that these higher order loops do not appear in either
Class I or Class II problems. 
If one takes the time derivative of eq.(\ref{eq:CformalIII}), 
one obtains, after straightforward but tedious calculations, the
following expression;
\begin{equation}
\diff{{\bf C}(t)}{t}
= -k_{\mbox{\scriptsize B}}T 
{\mbox{\boldmath$\chi$}}(t)
+\mbox{(Higher order loops)}.  
\label{eq:FDTIII}
\end{equation} 
Again, the higher order loops always 
contain at least one multiple of ${\cal V}^{(I)}{\cal V}^{(II)}$. 
One of the lowest order terms is an integral such as
\begin{equation}
\fl\hspace*{1.5cm}
\int\!\!{\mbox{d}} t'~
{\cal V}_{i\alpha\beta}{G}_{\alpha\lambda}(t-t')C_{\beta\mu}(t-t')
{\cal V}^{(I)}_{l\lambda\mu}
{\cal V}^{(II)}_{l\alpha'\beta'}
{G}^{\dagger}_{\alpha'\lambda'}(t')C_{\beta'\mu'}(t') 
{\cal V}_{j\lambda'\mu'}. 
\end{equation}
Note that this term is an irreducible loop in the field theoretic
language, which means that this can not be represented by any
simpler renormalized diagram. 
Since the original Langevin equation, eq.(\ref{eq:expansion}), itself 
{\it does} satisfy the FDT, the failure of the FDT in
eq.(\ref{eq:FDTIII}) is  
attributed to the inconsistencies of the loop expansion with the FDT. 
In other words, a naive loop expansion using the bare fields 
${\bf z}=({\bf x}, {\hat{\bf{x}}})$ for Class III problems does not
preserve the FDT at the each level of expansion; 
The higher order diagrams shown in eq.(\ref{eq:FDTIII})
are cancelled only by taking the next higher order loops in the MSR loop
expansion in Section 3.  

The failure to derive the FDT at the one-loop level for Class III 
leads to the failure of deriving a MCT type equation such as 
eq.(\ref{eq:mct}). 
By substituting eqs.(\ref{eq:CformalIII}) and (\ref{eq:FDTIII}) into 
the equation for $C_{ij}(t)$, eq.(\ref{eq:MCTC}), we obtain
\begin{equation}
\fl\hspace*{1.0cm}
\diff{C_{ij}(t)}{t}
= \mu_{i\alpha}C_{\alpha j}(t) +
\int_{0}^{t}\!\!{\mbox{d}} t_{1}~{E}_{i\alpha}(t-t_1)C_{\alpha j}(t_1)
+\mbox{(Higher order loops)},   
\label{eq:mctIII}
\end{equation}
where ${E}_{ij}(t)$ is given by the same expression as
eq.(\ref{eq:E-mct}) except that ${\cal V}$ is given by eq.(\ref{eq:Vijk3})
and 
${\cal L}$ in the second vertex, 
$\left( {\cal V}-2{\cal L} \right)$
in eq.(\ref{eq:E-mct}) is replaced by 
\begin{equation}
L^{(0)}_{i\alpha}\Lambda_{\alpha jk}
+ L^{(1)}_{i\alpha,j}\Omega_{\alpha k}
+ L^{(1)}_{i\alpha,k}\Omega_{\alpha j}. 
\end{equation}
The higher order loops in eq.(\ref{eq:mctIII}) 
are again irreducible diagrams which
do not appear if either ${\cal V}^{(I)}$ or
${\cal V}^{(II)}$ is absent. 
Eq.(\ref{eq:mctIII}) shows that the standard MCT equation for the Class
III is regarded as the {\it uncontrollable} approximation in the field
theoretic sense 
in that it neglects
terms which are present in the original set of 
equations, eqs.(\ref{eq:MCTC}) and (\ref{eq:MCTG}). 
The consequences of this are discussed below. 

\section{Discussion}

In order to see how the result shown in the previous sections 
are related to real fluids, let us consider the 
Langevin equation for a colloidal suspension given by eq.(\ref{eq:dean}).  
The Class II equation is derived by approximating the entropy given by 
eq.(\ref{eq:S0}) with its Gaussian form. 
Neglecting the terms of higher order than quadratic, one has
\begin{equation}
S \simeq  S_{0}-
\frac{k_{\mbox{\scriptsize B}}}{2}\int\!\!\mbox{d}{\bf k}~ 
\frac{|\delta\rho_{\bf k}|^2}{NS(k)},
\label{eq:Sappx}
\end{equation}
is the Fourier transform of $\delta\rho({\bf r})$,
$N$ is the total number of the particles, 
and 
$S(k)=N^{-1}\left\langle |\delta\rho_{\bf k}|^2 \right\rangle$ 
is the static structure factor. 
As mentioned in Section 3, this approximation has been shown to hold
over  a wide range of length scales in real
liquids\cite{Crooks1997,Hummer1996}. 
It is also compatible with the functionals used to derive integral
equations of fluid structure\cite{chandler1993}. 
It is not expected to hold on very short length scales where density
fluctuations are Poissonian and the ideal gas entropy is essential. 
On the other hand, density fluctuations on such length scales are not
expected physically to be effective in providing glassy behavior. 
It is plausible that the approximation eq.(\ref{eq:Sappx}) may be used
(along with an appropriate large wavevector cutoff) in the treatment of
realistic fluids. 
Indeed, the approximation, eq.(\ref{eq:Sappx}), is used in
Ref.\cite{schmitz1993}. 
Using eqs.(\ref{eq:Lij}) and (\ref{eq:Sappx}), the MCT equation,
eq.(\ref{eq:mct}), for the density correlation function  
$F(k,t)=N^{-1}\left\langle \delta\rho_{\bf k}(t)\delta\rho_{\bf -k}(0)
\right\rangle$ 
is written as 
\begin{equation}
\pdif{F(k, t)}{t}= -\frac{Dk^2}{S(k)}F(k,t) 
+ \int_{0}^{t}\!\!\mbox{d}t_1~  M'(k, t-t_1)F(k, t_1)
\label{eq:gotze1}
\end{equation}
with the memory kernel given by 
\begin{equation}
M'(k, t) = 
\frac{D^2k^2}{2\rho_{0}S(k)}\int\!\!\frac{{\mbox{d}}{\bf q}}{(2\pi)^3}
\left\{ \frac{\hat{\bf k}\cdot{\bf q}}{S(q)}
       +\frac{\hat{\bf k}\cdot{\bf p}}{S(p)}
\right\}^2 F(q, t)F(p, t),
\label{eq:memory1}
\end{equation}
where ${\bf p} = {\bf k}-{\bf q}$. 
Note that the vertex 
$\hat{\bf k}\cdot{\bf q}/S(q)+\hat{\bf k}\cdot{\bf p}/S(p)$ is precisely
the one that appears in Ref.\cite{schmitz1993}, where the Gaussian
approximation to the entropy is also made. 
This equation should be compared with the standard MCT
equation which has been derived using the projection operator method
with the decoupling approximation\cite{cichocki1987} where 
\footnote{Note that
eq.(\ref{eq:gotze1}) 
is different from the MCT equation used in the supercooled fluids;
$-Dk^2 F(k,t_1)$ appears in the place of 
$\partial F(k,t_1)/\partial
t_1$\cite{cichocki1987,franosch1997,kawasaki1995c}.   
The difference originates from the overdamped nature of the starting
diffusion equation, eq.(\ref{eq:dean}). 
One obtains the $\partial F(k,t_1)/\partial t_1$ term 
if one incorporates the momentum density as well as
the number density as stochastic variables. 
Technically, this is equivalent to using the irreducible projection
operator introduced by Cichocki {\it et al.}\cite{cichocki1987}.
}
\begin{equation}
\fl\hspace*{1.0cm}
M(k, t) = 
\frac{\rho_{0}D^2k^2}{2}\int\!\!\frac{{\mbox{d}}{\bf q}}{(2\pi)^3}
\left\{ \hat{\bf k}\cdot{\bf q}c(q)+\hat{\bf k}\cdot{\bf p}c(p)
\right\}^2 F(q, t)F(p, t).
\label{eq:memory2}
\end{equation}
We observe that there is a difference between 
eqs.(\ref{eq:memory1}) and (\ref{eq:memory2}):
$1/S(q)$ appears in the vertex function of eq.(\ref{eq:memory1}),
whereas the direct correlation function $\rho_{0}c(q)=1-1/S(q)$ shows up 
in eq.(\ref{eq:memory2}). 
This difference traces back to the Gaussian approximation,
eq.(\ref{eq:Sappx})\cite{das1996}.
The entropy of fluids is not Gaussian due primarily to 
ideal gas part, $\rho\ln\rho$, in eq.(\ref{eq:S0}). 
As discussed in Introduction, the nonlinearities of realistic
fluids come both from the entropy (in this case, $\rho\ln\rho$)
and the kinetic coefficient and
therefore realistic fluids are destined to belong to Class III over the 
entire range of wavevectors.   
Indeed, if the full expression for $S$, eq.(\ref{eq:S0}),
instead of approximated form of eq.(\ref{eq:Sappx}) is used, one sees
that the non-quadratic term of $S$ gives a vertex of the form of 
$-D(\hat{\bf k}\cdot{\bf q} +\hat{\bf k}\cdot{\bf p})$ which, combined
with
$D\{ \hat{\bf k}\cdot{\bf q}/S(q) +\hat{\bf k}\cdot{\bf p}/S(p)\}$, 
leads to 
$\rho_{0}D\{\hat{\bf k}\cdot{\bf q}c(q) +\hat{\bf k}\cdot{\bf p}c(p)\}$.
Using the full expression for $S$ means that the dynamics now belongs to 
Class III. 
As illustrated in Section 5, one obtains the memory kernel in the form
of eq.(\ref{eq:memory2}) but there are always extra terms which are 
a direct consequence of inconsistencies of the FDT with the one-loop
approximation in the MSR formalism, at least 
if the loop expansion is made directly with physical density modes as
field variables\footnote{Recent unpublished work by G. Biroli,
A. Lefevre, and J.-P. Bouchaud shows that if a transformed set of modes
is used, FDT can be recovered for the full class III problem, although,
at the time of submission of this work, other mathematical difficulties
arise that render the resulting equations unusable in the present
form.
In particular, the vertex that results is distinct from that of
eq.(\ref{eq:memory2}), and thus the one-loop approximation does not
yield the usual form of the standard idealized MCT of G\"{o}tze
and coworkers. 
}.   
This argument is true for arbitrary orders of the loop expansion.  
This conclusion 
implies that there is no simple systematic 
way to derive eq.(\ref{eq:memory2})
from eq.(\ref{eq:dean}) using  the standard field theoretic method. 
This inconsistency is not relevant as far as one is concerned only with
the equilibrium state, because one may always adopt an approximation 
where one neglects higher order terms in eq.(\ref{eq:mctIII}) 
and ``define'' the response function via the FDT instead of solving
the equation for ${\bf G}(t)$, eq.(\ref{eq:MCTG}), separately. 
But we can not do so if the system is in nonequilibrium state. 
It is desirable to develop such an expansion method that preserves the
FDT relation at each level of the perturbative expansion. 

Even with the difficulties discussed in this letter, 
the MSR or field theoretic approach is still an attractive route to
attack out-of-equilibrium supercooled fluids, in that 
it is systematic and one does not need to evaluate the nonequilibrium
measure which is required in alternative approaches such as the
projection operator technique. 
Models which belong to Class I have been already discussed extensively  
in the context of spin glasses and even for supercooled
fluids\cite{siggia1985}. 
However, 
it is difficult to construct realistic models of Class I which can 
incorporate the effect of
changes of structure embodied in $S(k)$ observed in
simulations of aging\cite{kob2000b} and sheared
systems\cite{miyazaki2004,fuchs2002}, which is argued to be essential
for the  
violation of the FDT of real fluids\cite{Szamel2004d}.    
The Class II system derived here is a better candidate 
as a realistic ``model fluid''. 
As discussed above,
the Gaussian approximation for the entropy is known to be a good
description for wide ranges of densities and
length scales \cite{Crooks1997,Hummer1996}. 
The equation for equal-time correlation functions such as 
$S(k)$ should be constructed in the same manner as  
the MCT equation derived here. 
The solution could be plugged into the vertex functions of the set of 
the MCT equations, eqs.(\ref{eq:MCTC}) and (\ref{eq:MCTG}). 
Eventually, these three equations can be solved self-consistently.
Performing such a calculation will require 
some consideration of a wavevector cutoff that would eliminate spurious
divergence that arise 
from the {\it approximate} vertex functions (eq.(\ref{eq:memory1})). 
Calculations in this direction are underway.

In summary, in the present work, we have broadened the range of
applicability  
of the MSR approach by extending the method 
to dynamical processes with the multiplicative noise. 
This is a necessary step in the treatment of the Brownian dynamics of
colloidal suspensions. 
This formalism still does not cover real fluids over all length scales 
and 
it is found that there is no direct compatibility 
between the MCT equations derived from MSR approach
and from the the projection operator method.  
We proposed an approximate but feasible method 
to explore nonequilibrium supercooled fluids using the formalism
discussed in this paper.
The MSR method is not restricted to evaluation of the two point
correlation functions nor to the lowest order loop expansion. 
Extension of the method to the multipoint correlation functions and to 
higher order 
loops will be essential for understanding 
growing length scales which are hidden in the supercooled
fluids\cite{biroli2004}.  
The formulation presented in this paper
will serve as first step towards such extensions.

\ack
The authors acknowledge support from NSF grant \#0134969.  
We would like to thank G. Biroli, M. E. Cates, D. S. Dean,
P. L. Geissler, A. Onuki, S. Ramaswamy, and A. Yoshimori for helpful
discussions.


\section*{References}


\begin{thebibliography}{10}

\bibitem{gotze1992}
G{\"{o}}tze W and Sj{\"{o}}gren L
1992
{\it Rep. Prog. Phys.}
{\bf 55}
241

\bibitem{gotze1999}
G{\"{o}}tze W
1999
\newblock {\it J. Phys: Condens. Matter}, 
{\bf 11}
A1

\bibitem{biroli2004}
Biroli G and Bouchaud J-P
2004
{\it Europhys. Lett.}
{\bf 67}
21;

\bibitem{das1986}
Das S P and Mazenko G F
1986
{\it Phys. Rev.} {\rm A} 
{\bf 34}
2265

\bibitem{ediger2000}
Ediger M D 
2000
\newblock {\it Ann. Rev. Phys. Chem.}
{\bf 51}
99

\bibitem{yamamoto1998c}
Yamamoto R and Onuki A
1998
\newblock {\it Phys. Rev. {\rm E}}
{\bf 58}
3515

\bibitem{kob2000b}
Barrat J -L and Kob W
1999
{\it Europhys. Lett.}
{\bf 46}
637; 
Kob W, Barrat J -L, Sciortino F, and Tartaglia P 
2000
{\it J. Phys: Condens. Matter}
{\bf 12}
6385

\bibitem{berthier2002b}
Berthier L, Barrat J -L, and Kurchan J
2000
{\it Phys. Rev.} {\rm E}
{\bf 61} 
5464;
Barrat J L and Berthier L 
2000
{\it ibid}
{\bf 63}
012503

\bibitem{crisanti2003}
Crisanti A and Ritort F
2003
{\it J. Phys.} {\rm A} 
{\bf 36}
R181

\bibitem{martin1973}
Martin P C, Siggia E D, and Rose H A 1973
{\it Phys. Rev.} {\rm A}
{\bf 8} 
423

\bibitem{McComb2004}
McComb W D 2004
{\it Renormalization methods: A guide for beginners}
(Oxford, Oxford University Press)

\bibitem{bouchaud1996}
Bouchaud J -P, Cugliandolo L, Kurchan J, and M\'{e}zard M 
1996
{\it Physica {\rm A}} 
{\bf 226}
243

\bibitem{schmitz1993}
Schmitz R, Dufty J W, and De P 
1993
{\it Phys. Rev. Lett.} 
{\bf 71}
2066

\bibitem{kawasaki1997e}
Kawasaki K and Miyazima S 
1997
{\it Z. Phys. {\rm B}: Condens. Matter}
{\bf 103}
423

\bibitem{dean1996}
Dean D S 1996
{\it J. Phys.} {\rm A} 
{\bf 29}
L613

\bibitem{kawasaki1994}
Kawasaki K
1994
{\it Physica {\rm A}}
{\bf 208}
35

\bibitem{Archer2004}
Archer A J and Rauscher M
2004
{\it J. Phys.} {\rm A} 
{\bf 37}
9325


\bibitem{cichocki1987}
Cichocki B and Hess W
1987
\newblock {\it Physica {\rm A}}
{\bf 141}
475

\bibitem{szamel1991}
Szamel G and L{\"{o}}wen H
1991
{\it Phys. Rev.} {\rm A}
{\bf 44}
8215

\bibitem{franosch1997}
Franosch T, Fuchs M, G{\"{o}}tze W, Mayr M R, and Singh A P
1997 
{\it Phys. Rev.} {\rm E}
{\bf 55}
7153

\bibitem{deker1975}
Deker U and Haake F 
1975
{\it Phys. Rev.} {\rm A}
{\bf 11}
2043

\bibitem{vankampen1981}
van Kampen N G 1981
{\it Stochastic Processes in Physics and Chemistry}
(Amsterdam: North-Holland)

\bibitem{landau1959}
Landau L D and Lifshitz E M
1959
\newblock {\it {"{Fluid Mechanics}"}}.
(New York, Pergamon)

\bibitem{Miyazaki1996}
Miyazaki K, Kitahara K, and Bedeaux D
1996
\newblock {\it Physica {\rm A}}
{\bf 230}
600

\bibitem{graham1973}
Graham R
1973
{\it Springer tracts in modern physics}
ed  G H{\"o}hler
(Berlin: Springer),
{\bf 66}
1

\bibitem{kawasaki1970}
Kawasaki K 
1970
{\it Ann. Phys.} {\bf 61}
1

\bibitem{deker1979}
Deker U 1979
{\it Phys. Rev.} {\rm A}
{\bf 19} 
846

\bibitem{phythian1977}
Phythian R 
1977
{\it J. Phys.} {\rm A} 
{\bf 10}
777

\bibitem{jensen1981}
Jensen R V 
1981
{\it J. Stat. Phys.} 
{\bf 25}
183

\bibitem{Crooks1997}
Crooks G E and Chandler D
1997
{\it Phys. Rev.} {\rm E}
{\bf 56}
4217

\bibitem{Hummer1996}
Hummer G, Garde S, Garc{\'{i}}a A E, Pohorille A, and Pratt L R
1996
{\it Proc. Natl Acad. Sci.}
{\bf 93}
8951

\bibitem{chandler1993}
Chandler D
1993
{\it Phys. Rev.} {\rm E}
{\bf 48}
2898

\bibitem{kawasaki1995c}
Kawasaki K
1995
{\it Physica {\rm A}}
{\bf 215}
61

\bibitem{das1996}
Das S P
1996
{\it Phys. Rev.} {\rm E}
{\bf 54}
1715

\bibitem{siggia1985}
Siggia E
1985
{\it Phys. Rev.} {\rm A}  
{\bf 32}
3135

\bibitem{miyazaki2004}
Miyazaki K and Reichman D R 
2002
{\it Phys. Rev.} {\rm E} 
{\bf 66}
050501(R);
Miyazaki K, Reichman D R, and Yamamoto R 
2004
{\it Phys. Rev.} {\rm E} 
{\bf 70}
011501

\bibitem{fuchs2002}
Fuchs M and Cates M E 
2002
{\it Phys. Rev. Lett. } 
{\bf 89}
248304;
Fuchs M and Cates M E     
2003
{\it Faraday Discuss.} 
{\bf 123}
267

\bibitem{Szamel2004d}
Szamel G
2004
{\it Phys. Rev. Lett.}
{\bf 93}
178301

\end{thebibliography}
\end{document}